\documentclass[conference]{IEEEtran}
\IEEEoverridecommandlockouts
\usepackage{forest}
\usepackage{xcolor}
\usepackage[utf8]{inputenc} 
\usepackage[T1]{fontenc}
          
\usepackage{booktabs}       
\usepackage{amsfonts}       
\usepackage{nicefrac}     
\usepackage{microtype}

\usepackage{lipsum}
\usepackage{fancyhdr}      
\usepackage{graphicx}       
\graphicspath{{media/}}
\usepackage{array}
\usepackage{cite}
\usepackage{amsmath}
\let\labelindent\relax
\usepackage{enumitem}

\usepackage{multirow}

\usepackage{arydshln}

\usepackage{makecell}

\usepackage[hidelinks,breaklinks=true]{hyperref}
\usepackage{xurl}
\usepackage{float}

\newlist{rqs}{enumerate}{1}
\setlist[rqs]{label*=\textbf{RQ\arabic*}}

\def\IEEEkeywordsname{Keywords}

\begin{document}

\title{LeafTutor: An AI Agent for Programming Assignment Tutoring}

\author{\IEEEauthorblockN{Madison Bochard\IEEEauthorrefmark{1}, Tim Conser\IEEEauthorrefmark{1}, Alyssa Duran\IEEEauthorrefmark{1}, Lazaro Martull\IEEEauthorrefmark{1}, Pu Tian\IEEEauthorrefmark{2}, and Yalong Wu\IEEEauthorrefmark{1}}
\IEEEauthorblockA{\IEEEauthorblockA{\IEEEauthorrefmark{1}Department of Computing Sciences, University of Houston--Clear Lake, TX, USA\\ Emails: \{bochardm3889, consert1575, durana7167, martullgonzal6174, wuy\}@uhcl.edu\\
\IEEEauthorrefmark{2}Computer Science Program, Stockton University, NJ, USA. Email: pu.tian@stockton.edu}}\thanks{Work in progress.}}

\maketitle

\begin{abstract}
High enrollment in STEM-related degree programs has created increasing demand for scalable tutoring support, as universities experience a shortage of qualified instructors and teaching assistants (TAs). To address this challenge, LeafTutor, an AI tutoring agent powered by large language models (LLMs), was developed to provide step-by-step guidance for students. LeafTutor was evaluated through real programming assignments. The results indicate that the system can deliver step-by-step programming guidance comparable to human tutors. This work demonstrates the potential of LLM-driven tutoring solutions to enhance and personalize learning in STEM education. If any reader is interested in collaboration with our team to improve or test LeafTutor, please contact Pu Tian (pu.tian@stockton.edu) or Yalong Wu (wuy@uhcl.edu).
\end{abstract}


\section{Introduction}
\label{sec:introduction}

Enrollment in STEM-related degree programs has risen steadily. For example, computer science and engineering (CSE)-related bachelor’s degrees in the 2023–2024 academic year reached an enrollment rate of 6.8 percent, nearing pre-pandemic levels~\cite{batten2025enrollment}. This sustained growth has intensified long-standing challenges in higher education: a limited number of qualified instructors and teaching assistants (TAs) must now support increasingly large course sections, limiting the timeliness and individualization of student feedback. As class sizes expand while instructional resources remain stagnant, scalable solutions for individualized support have become essential, especially for programming courses where students benefit from step-by-step guidance. 

An AI agent is an autonomous intelligent system capable of scalable solutions for automating tasks, solving complex problems, and engaging in conversational assistance. By maintaining contextual memory of prior interactions, such agents can generate personalized responses tailored to user expectations~\cite{gutowska2025aiagents}. These characteristics of scalability and personalization align closely with the instructional demands of expanding STEM programs. However, although university students presently rely on AI tools such as OpenAI ChatGPT, Microsoft Copilot, and Google Gemini for homework assistance, most existing systems emphasize delivering direct answers rather than providing the guided, step-by-step feedback typical of human tutoring.

To bridge this gap, we develop LeafTutor, an AI tutoring agent designed to deliver guided programming assistance rather than complete answers. LeafTutor combines large language model (LLM) reasoning with assignment-specific retrieval alongside code execution feedback to emulate human tutoring behavior. Our main contributions are as follows:

\begin{itemize}
    \item We demonstrate that LLMs not only enhance coding quality for software developers but also provide interactive tutoring for entry-level programming courses.
    \item The system features two integrated modules: (i) an instructor interface for uploading assignment instructions and lecture materials, and (ii) a student interface that provides AI-driven conversational assistance, file upload support for class materials and code submissions, and an interactive code editor with a backend compiler.
    \item We evaluate LeafTutor through internal testing using real programming assignments to assess tutoring behavior, functional performance, and potential for classroom deployment.
\end{itemize}

\section{Related Works}
\label{sec:relatedwork}

LLMs have significantly advanced the way information is processed and explained. Their ability to reason over complex inputs and generate context-aware responses has driven innovation across many domains such as writing, research, and software development. Recent evaluations highlight that large-scale models can perform multi-step reasoning and offer contextual feedback across diverse tasks \cite{openai2023gpt4}. These are the qualities that make LLMs well suited for addressing challenges in programming education, where students often need step-by-step explanations and adaptive guidance rather than direct solutions.

In education, AI is being explored as a supportive learning companion that promotes understanding through dialogue. Research on \textit{Generative Agents} \cite{park2023generativeagents} showed that autonomous systems can model memory, reasoning, and interaction patterns similar to human tutors, suggesting how AI might personalize instruction. Several AI-powered tools already assist programming learners. \textit{PythonTutor} \cite{pythontutor} visualizes code execution but offers limited conceptual feedback, while \textit{CodingZap} \cite{codingzap} provides automated help but often yields complete solutions. Other tools such as \textit{CodeBuddy} \cite{codebuddy} and IDE assistants powered by \textit{Codex} \cite{codex} demonstrate natural language support within development environments, although they typically function as stand-alone aids with limited educational structure.

Building on these developments, \textit{LeafTutor} applies an LLM-based conversational agent specifically for programming assignments. The GPT-4 Technical Report \cite{openai2023gpt4} and Google's Gemini project \cite{google2024gemini} illustrate the technical foundation that enables this type of adaptive tutoring, even though neither of them directly target education. Unlike existing code helpers, \textit{LeafTutor} guides learners through iterative steps—clarifying logic, identifying syntax issues, and encouraging problem-solving. While the current system operates as a standalone prototype, its design emphasizes conceptual understanding and responsible AI-assisted learning, setting the base or groundwork for potential integration into classroom environments.
  
\section{System Architecture}\label{sec:architecture}
  
\begin{figure}[!b]
  \centering
  \includegraphics[scale=0.8]{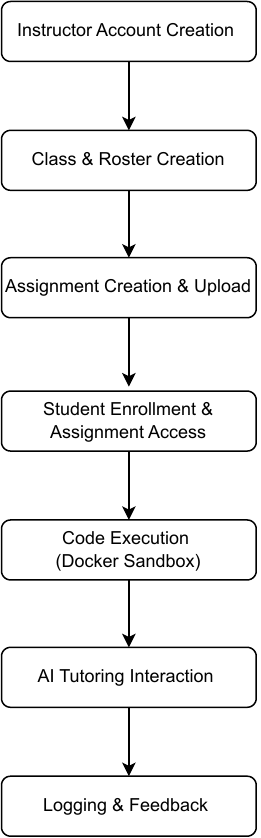}
  \caption{Overall workflow of the \textit{LeafTutor} system, from instructor account creation and material ingestion to student assignment interaction and AI-assisted feedback.}
  \label{fig:pipeline}
\end{figure}

\begin{figure*}[htbp]
    \centering
    \includegraphics[scale=0.75]{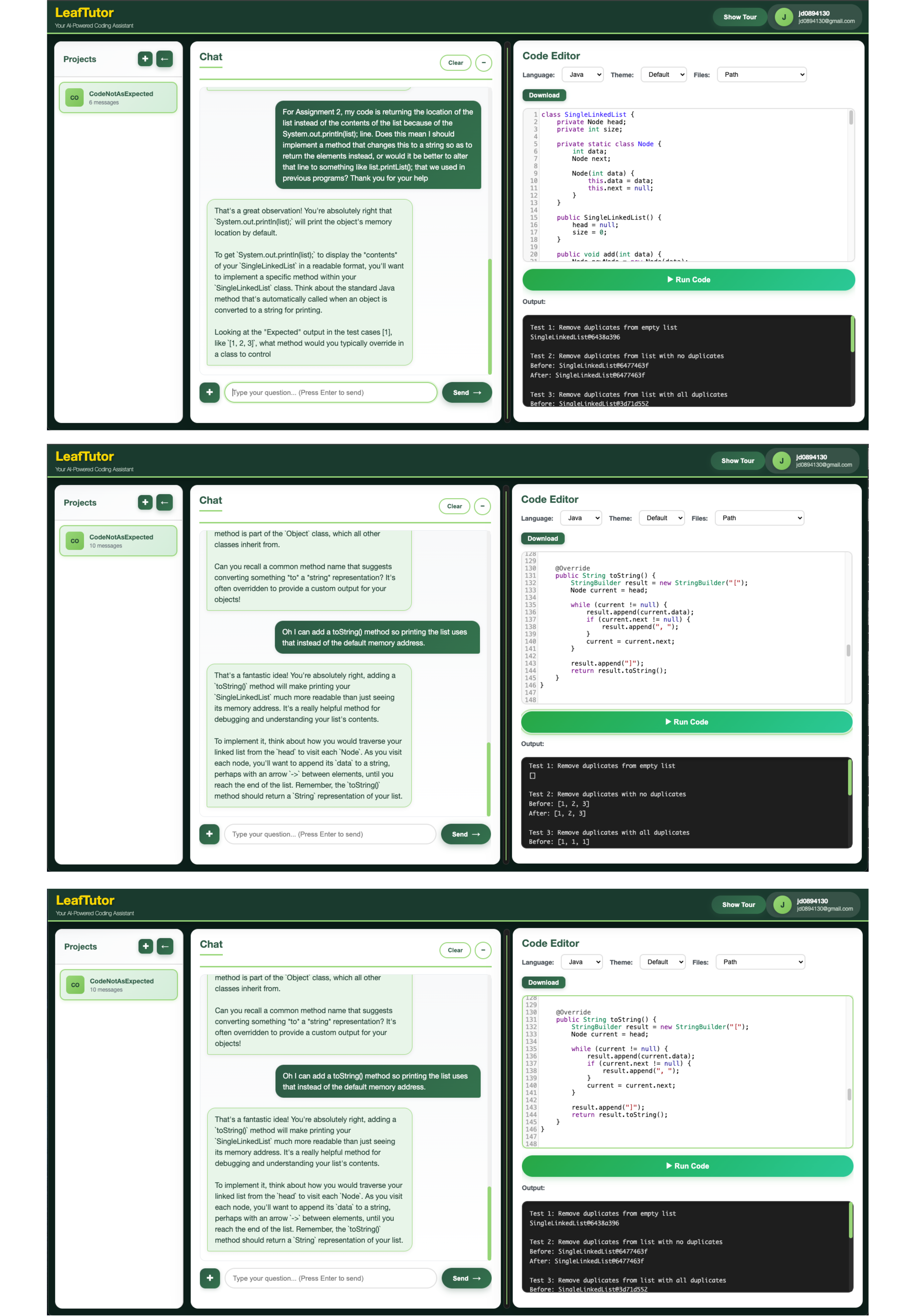}
    \caption{Code Not As Expected Error. Interaction demonstrating LeafTutor diagnosing a non-error output issue. After detecting that the student’s linked-list implementation prints a memory reference rather than its contents, the tutor explains the cause, recommends implementing a toString() method, and confirms the corrected behavior once added.}
    \label{fig:student1}
\end{figure*}

\begin{figure*}[htbp]
    \centering
    \includegraphics[scale=0.75]{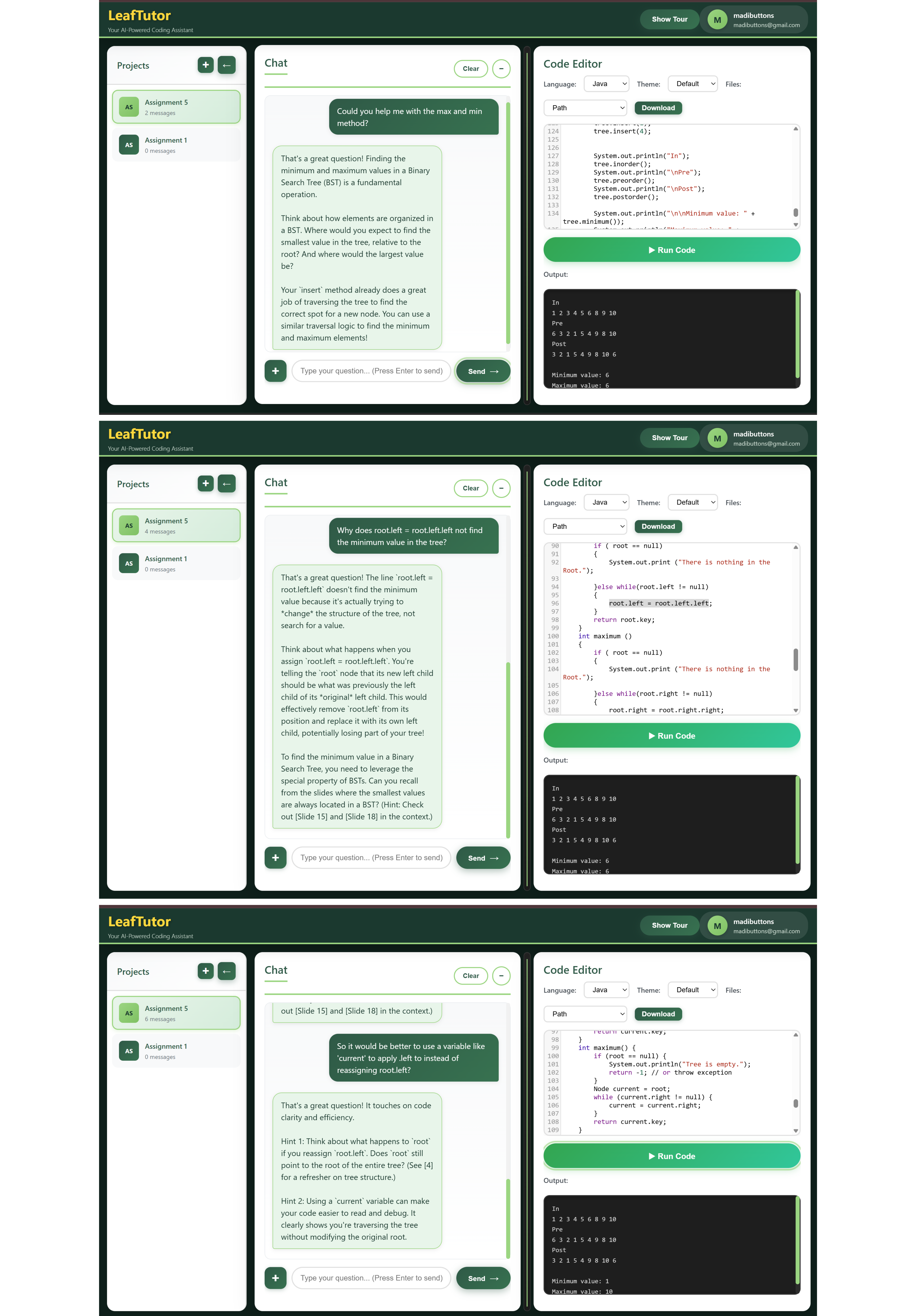}
    \caption{Logical Error. LeafTutor identifying and correcting a logical error in BST minimum/maximum value computation.}
    \label{fig:student2}
\end{figure*}

FIGURE~\ref{fig:pipeline} illustrates the workflow of the LeafTutor system. The platform emulates the experience of a teaching assistant consultation during office hours.

To start, an instructor sets up an account to manage courses, class rosters, and assignments. 
For each assignment, the instructor uploads instructions, sample solutions, course materials, and remarks. They are utilized as preliminary input when a student asks questions regarding an assignment. 

When there is a need for tutoring, enrolled students upload their incomplete work and interact with the AI agent to receive guidance on assignment completion. Students can inquire about assignment requirements, implementation strategies, and problem-solving approaches, with the AI agent providing contextual assistance and pedagogical guidance. The platform includes an integrated code editor to facilitate completion of coding assignments.

The backend offers compilers to run the uploaded code. When compilation or runtime errors occur within the code editor, students can query the integrated AI agent for diagnostic assistance. Students submit inquiries describing the problem or requesting clarification regarding the error message. The AI agent analyzes the error context and provides explanatory feedback to guide the student toward resolution, facilitating independent problem-solving and learning reinforcement.

The AI interaction layer internally invokes the LLM API to support the tutoring process. Consequently, when a student poses a query such as “What should I do for this assignment?”, the system identifies relevant instructional content and supplies it to the AI model; this ensures that responses remain grounded in assignment-specific information rather than generic advice. 

Each tutoring request is saved, which provides a structured dataset for longitudinal analysis of learning behaviors and the effectiveness of AI-assisted tutoring.

Collectively, these components form an integrated pipeline wherein instructor-provided materials are processed, embedded, and stored, and subsequently linked to student activities. This integration enables context-aware, real-time AI assistance that supports individualized learning and continuous assessment.

\section{Experimental Testing and Validation}
\label{sec:experiment}

To evaluate LeafTutor, four Java assignments were used to simulate student tutoring. For each evaluation, the development team uploaded instructor materials, such as an assignment document, sample solution code, and lecture materials, through the instructor interface. These materials were then retrieved by the AI tutor during student–side interactions. The system was evaluated across four categories: incorrect output, code-detail errors, logical errors, and runtime errors. FIGURE~\ref{fig:student1} and FIGURE~\ref{fig:student2} demonstrate the interfaces.

\subsection{Code Not As Expected Error}

This test assessed the AI tutor's ability to handle cases where the code runs but produces incorrect output. The assignment required implementing a \texttt{removeDuplicates()} method in a \texttt{SingleLinkedList} class and printing results consistent with the sample output from the assignment document. In the incorrect version, \texttt{System.out.println(list)} displayed the object’s memory reference rather than its contents. When prompted with “For Assignment 2, my code is returning the location of the list instead of the contents of the list because of the \texttt{System.out.println(list)} line…”, the AI tutor correctly identified that \texttt{System.out.println(list)} invoked the default \texttt{Object.toString()} method and advised that overriding the \texttt{toString()} method was necessary to generate a readable list representation. After updating the code, the output matched the expected results. This demonstrated the tutor’s ability to detect misleading output in otherwise functional code and provide targeted guidance to meet assignment specifications.

\subsection{Code-Detail Error}

To test fine-grained error detection, an assignment required implementing a \texttt{Shape} interface and applying the correct geometric formulas in each shape class was used, with the \texttt{Triangle} class intentionally returning only the semi-perimeter portion of Heron’s area formula. When asked “What’s the problem with my triangle area?”, the AI tutor compared the submitted implementation with the assignment requirements and correctly identified the missing step of Heron's formula. It referenced the assignment instructions and explained why the returned value was incomplete, enabling the tester to update the method accordingly. This showed that the tutor can recognize subtle implementation mistakes and produce targeted, assignment-aligned feedback for correcting them.

\subsection{Logical Error}

The third test evaluated the tutor’s ability to diagnose errors in algorithmic reasoning. The assignment required traversals for a binary search tree (BST) and methods to return the smallest and largest values. In the intentionally flawed version of the code, the \texttt{minimum()} and \texttt{maximum()} methods reassigned \texttt{root.left} to \texttt{root.left.left} and \texttt{root.right} to \texttt{root.right.right}, inadvertently discarding nodes and preventing correct results. When initially prompted with “Could you help me with the max and min methods?”, the tutor asked clarifying questions to guide reasoning and referenced structurally similar logic already present in the code. Follow-up prompts led it to explain that reassigning \texttt{root.left} and \texttt{root.right} destroys nodes rather than traversing the tree. It referenced lecture materials and confirmed,  after an additional prompt, that using a temporary traversal variable better aligned with standard BST logic. This demonstrated the tutor’s ability to diagnose structural logic errors and use contextual materials to support correction.

\subsection{Runtime Error}

The final test examined the tutor’s handling of runtime failures, specifically an ``Array Index Out Of Bounds Exception" message in an array-based binary tree implementation. Unlike the previous test assignments, this one originated from an in-class quiz, thus the instructor-provided materials consisted only of the solution code and lacked any formal assignment instructions for the AI tutor to reference. In the incorrect version, methods such as \texttt{left(int i)}, \texttt{right(int i)}, and \texttt{setLeft(int i, int value)} used index calculations or boundary checks that did not consistently prevent access to invalid array positions, resulting in a runtime error. When prompted with “Index out of bound, but I cannot move on,” the AI tutor correctly recognized that the failure stemmed from incorrect index calculations and provided high-level guidance on reviewing child index formulas. However, it struggled to pinpoint the exact faulty expression and often offered general suggestions rather than precise fault localization. This highlighted a current limitation in its ability to trace and isolate runtime issues, especially when detailed assignment instructions are not provided.

\subsection{Results}

Across the first three assignments, LeafTutor consistently interpreted user prompts, utilized uploaded materials, and provided corrections aligned with assignment expectations. It effectively handled conceptual, structural, and implementation errors. In contrast, the runtime-error test highlighted limitations in pinpointing the exact faulty expression, even though the tutor recognized the general source of the failure. Overall, the system performs well on structured tasks with clear requirements but would benefit from enhanced error-trace analysis and finer-grained debugging support for unstructured or minimally guided code.

\section{Future Work}
\label{sec:futurework}

Looking ahead, the continued development of LeafTutor is guided by a vision of transforming the way students and instructors interact with programming education. We imagine a learning environment in which students transition seamlessly between their university learning management systems and LeafTutor without experiencing workflow disruption. 

The tutoring experience itself is envisioned to become more adaptive and personalized. Rather than focusing only on assignment-specific context, future versions of LeafTutor could build a persistent learning profile for each student, allowing the system to remember prior conversations, identify long-term misconceptions, and adjust explanations to match individual learning styles. As students encounter errors or seek guidance, LeafTutor could intelligently recommend curated documentation, libraries, or language-specific references to encourage independent skill development. Improvements to the interface may also include direct access to all instructor-provided instructional materials in their original formats, displayed alongside the chat and code panels to facilitate a more natural learning workflow.

Together, these future directions outline a path toward a more integrated, intelligent, and scalable tutoring ecosystem—one capable of evolving with the rapidly changing landscape of STEM education and providing learners with support that is adaptive, contextual, and deeply human-centered.

\section{Conclusion}\label{sec:done}
This work presented LeafTutor, an AI-driven tutoring system developed to address the growing need for scalable, individualized support in programming education amid rising enrollment and limited TA availability. Unlike existing AI tools that focus on providing direct answers, LeafTutor delivers step-by-step, context-aware guidance through an integrated chat and code-execution environment. Internal testing with real programming assignments demonstrated that LeafTutor produces adaptive and pedagogically aligned feedback, validating its potential for classroom integration. While these results highlight the feasibility of incorporating LLM-based tutoring into STEM curricula, further evaluation with external users is needed to assess learning outcomes across diverse course settings. Moving forward, expanded language support, large-scale user testing, and enhancements that adapt to a student's current level of understanding will advance LeafTutor toward providing personalized, accessible assistance that meets learners where they are in their learning journey. Overall, this work contributes to the advancement of LLM-driven educational tools aimed at enhancing accessibility and personalization in STEM education.


\bibliographystyle{IEEEtran}
\bibliography{ref}

\end{document}